\documentclass[sigconf]{acmart}

\AtBeginDocument{%
  \providecommand\BibTeX{{%
    \normalfont B\kern-0.5em{\scshape i\kern-0.25em b}\kern-0.8em\TeX}}}

\setcopyright{acmlicensed}
\copyrightyear{2023}
\acmYear{2023}
\acmDOI{XXXXXXX.XXXXXXX}

\acmConference[CHI EA '23]{}{April 23–28, 2023}{Hamburg, Germany}
\makeatletter
\renewcommand{\@copyrightowner}{© Authors | ACM 2023. This is the author's version of the work. It is posted here for your personal use. Not for redistribution. The definitive Version of Record was published in the CHI'23 proceedings, http://dx.doi.org/10.1145/3544549.3585830.}
\makeatother

\begin{document}

\title{Establishing Awareness through Pointing Gestures during Collaborative Decision-Making in a Wall-Display Environment}

\author{Valérie Maquil}
\affiliation{%
  \institution{Luxembourg Institute of Science and Technology}
   \country{Luxembourg}
}
\author{Dimitra Anastasiou}
\affiliation{%
	\institution{Luxembourg Institute of Science and Technology}
	\country{Luxembourg}
}

\author{Hoorieh Afkari}
\affiliation{%
	\institution{Luxembourg Institute of Science and Technology}
	\country{Luxembourg}
}

\author{Adrien Coppens}
\affiliation{%
	\institution{Luxembourg Institute of Science and Technology}
	\country{Luxembourg}
}

\author{Johannes Hermen}
\affiliation{%
	\institution{Luxembourg Institute of Science and Technology}
	\country{Luxembourg}
}

\author{Lou Schwartz}
\affiliation{%
	\institution{Luxembourg Institute of Science and Technology}
	\country{Luxembourg}
}

\renewcommand{\shortauthors}{Maquil et al.}

\begin{abstract}
  Sharing a physical environment, such as that of a wall-display, facilitates gaining awareness of others’ actions and intentions, thereby bringing benefits for collaboration. Previous studies have provided first insights on awareness in the context of tabletops or smaller vertical displays. This paper seeks to advance the current understanding on how users share awareness information in wall-display environments and focusses on mid-air pointing gestures as a foundational part of communication. We present a scenario dealing with the organization of medical supply chains in crisis situations, and report on the results of a user study with 24 users, split into 6 groups of 4, performing several tasks. We investigate pointing gestures and identify three subtypes used as awareness cues during face-to-face collaboration: narrative pointing, loose pointing, and sharp pointing. Our observations show that reliance on gesture subtypes varies across participants and groups, and that sometimes vague pointing is sufficient to support verbal negotiations.
\end{abstract}



\begin{teaserfigure}
\centering
  \includegraphics[width=.8\textwidth]{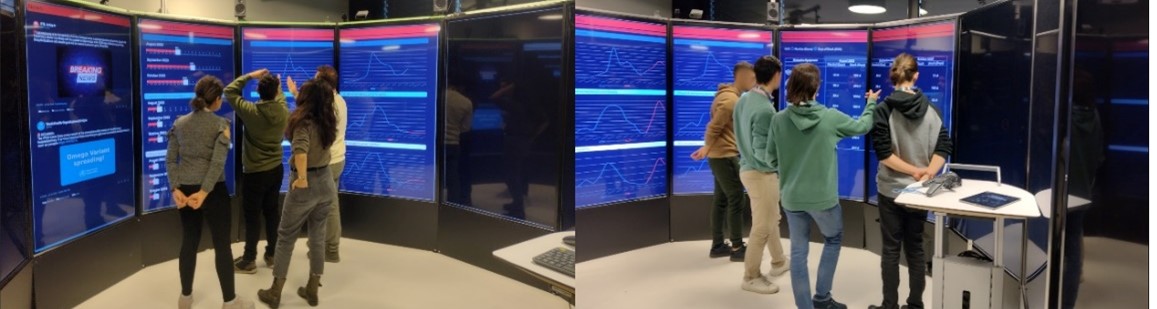}
  \caption{Groups interacting at a wall-display and frequently using mid-air pointing gestures to share awareness information.}
  \Description{}
  \label{fig:teaser}
\end{teaserfigure}


\maketitle

\section{Introduction}

Large interactive displays are increasingly placed in public locations, including museums, shops, airports, city settings, and offices. They enable large amounts of data to be presented in different scales and views, and support users to better identify details and gain more insights about the data \cite{rajabiyazdi2015understanding}. They therefore provide cognitive and perceptual benefits, and in general enhance user performance and satisfaction for such tasks (e.g., \cite{reda2015effects}). Another benefit of wall displays is that they inherently support (collocated) collaboration, as multiple users can access and view content at the same time, and also observe each other’s interactions. With this information they can coordinate their activities, anticipate others’ actions, and assist each other. The awareness information makes collaborators’ actions and intentions clear and allows them to seamlessly align and integrate their activities with those of other group members.
Existing prior work on collaborative technologies generally acknowledges the importance of a maintaining high awareness of the workspace, and various technological approaches have been proposed to support this awareness, e.g., when personal or separated workspaces are included. However, to date, there are only a few empirical or observational studies on how awareness is created and shared by users of large interactive displays \cite{hornecker2008collaboration, pinelle2008effects, anastasiou2020you}.
In this paper, we seek to study how mid-air pointing gestures are used by groups of four collocated participants to share awareness information during collaborative decision-making in a wall-display environment (see Figure \ref{fig:teaser}). We understand awareness information to be information about ongoing activities by users of the interactive wall displays, which may or may not involve workspace artefacts. The work is part of the research project ReSurf, \cite{resurf} aiming to enhance mixed-presence collaborative decision-making at physically distributed wall displays by adding a series of digital cues that mediate awareness information. To develop this support, we first study awareness information during collocated collaborative use and then use the insights gained to design awareness support for mixed-presence settings.
This paper is structured as follows: we first report on related work and then describe the design of our user study, including the collaborative decision-making application we have developed and instantiated in a wall-display environment. We then present our results on mid-air pointing gestures, focusing on their subtypes and their frequencies.

\section{Related work}
Interactive wall-displays have been used in research for various data intensive tasks, such as information visualization \cite{langner2018multiple}, visual analytics \cite{reda2015effects} or sensemaking \cite{andrews2010space}. Wall-sized displays have three common applications: (1) navigating a single, very large object (e.g. a complicated 3D molecule or an extremely large and detailed image), (2) comparing a large quantity of related visual content (such as photographs), and (3) juxta positioning data from different sources (e.g., photographs, video clips, articles, raw data tables, notes) \cite{beaudouin2012multisurface}. Interactive wall-displays were found to support several collaboration styles \cite{jakobsen2014up} as well as fluid transitions between personal and shared workspaces \cite{prouzeau2018awareness}. Because wall-displays provide a shared space, they support high awareness of other users’ actions and intentions \cite{yuill2012mechanisms}.
There are two most common types of awareness. Situation awareness refers to the acquisition and interpretation of information from the environment that is fundamental for subsequent decision making \cite{endsley1995measurement}. In this paper, we focus on workspace awareness, defined as the notion of monitoring the activity of others, which provides context for one’s activities \cite{dourish1992awareness}, \cite{gutwin2002descriptive}. It comprises knowledge about where others are working, what they are doing, and what they are going to do next. Awareness information is crucial for many of the activities related to collaboration: for coordinating actions, managing coupling, talking about the task, anticipating others’ actions, and finding opportunities to assist one another. It allows users to seamlessly align and integrate activities with those of other group members. The main sources of awareness information are i) people’s bodies, ii) workspace artefacts, and iii) conversation and gestures \cite{gutwin2002descriptive}. Previous research has shown that in a remote setting, individuals have faced difficulties in finding people and clarifying other participants’ thoughts and consequently, do not have common orientations and reference points \cite{herbsleb2000distance}, \cite{mark2002conventions}. Therefore, lack of awareness of other team members´ working processes is one of the drawbacks that a team may face while attempting to collaborate on a shared task, especially in a distant collaboration context. To address this problem emphasized by remote scenarios, prior work has implemented and evaluated various techniques ranging from simple remote cursors \cite{marrinan2017mixed} and video feeds \cite{kuechler2010collaboard} recorded from a standard camera, to digital embodiments such as virtual arms \cite{yamashita2011improving}, and even robot avatars \cite{oyekoya2013supporting}.
Only a few works have studied behaviour related to awareness around large interactive surfaces (tabletops and wall-displays): Hornecker et al. \cite{hornecker2008collaboration} explored awareness around tabletop displays through negative and positive awareness indicators. They identified i) interference (e.g., reaching for same object) and ii) verbal monitoring (“what did you do there?”) as negative awareness indicators, and i) reaction without explicit request, ii) parallel work on same activity without verbal coordination, among others, as positive awareness indicators. Building on this work, Tong et al. \cite{tong2017horizontal} compared horizontal vs. vertical surfaces with regards to collaboration and found out that in both conditions a good level of awareness could be maintained. However, in the vertical condition participants needed to do more explicit awareness work in the form of verbal shadowing. Pinelle et al. \cite{pinelle2008effects} studied direct touch and relative input techniques with different levels of embodiment for tabletop systems and showed that direct touch generally generated higher levels of awareness of other users’ location or activity, but that virtual embodiments were preferred by the majority of participants. More recently, Anastasiou et al. \cite{anastasiou2020you} studied how pointing gestures were used in low awareness situations at tabletop interfaces.
As far as pointing is concerned, it has been examined by many scholars, such as linguists, semioticians, psychologists, anthropologists and primatologists. The prototypical pointing gesture is a communicative body movement that projects a vector from a body part, with this vector indicating a certain direction, location, or object \cite{kita2003pointing}. Many scholars have examined pointing in various languages and cultures, e.g., Kendon and Versante \cite{kendon2003pointing} about variations of pointing gestured in Neapolitan and Wilkins \cite{wilkins2003pointing} who examined cultural variation in pointing gestures and discussed what this reveals about cross-cultural differences in the semiotics of pointing. As far as gestures in wall-displays are concerned, they were researched in the field of arranging data items \cite{liu2017coreach}, in terms of intuitiveness and effectiveness \cite{hespanhol2012investigating}, and how accurately they can refer to a target when observed on a video stream \cite{avellino2015accuracy}.
To sum up, while previous work has provided first insights about workspace awareness on large interactive displays, there is little knowledge on the role of mid-air pointing gestures in this context. In this paper, we seek to understand how mid-hair pointing gestures are used around wall-sized surfaces and how they support the sharing of awareness information with the aim of investigating how they can be replicated remotely on another display.

\begin{table*}
  \caption{The four tasks of the decision-making scenario and the types of data they include.}
  \label{tab:commands}
 \begin{tabular}{p{3.5cm}p{6.45cm}p{3.5cm}p{2.6cm}}
 \toprule
Task & Description & Data & Predominant roles\\ \midrule
1:   Estimate future COVID numbers  & Users need to read the news, estimate hospital and ICU occupancy, and adapt the   sliders according to their estimation.& Line graphs, Text  & Head of ICU \\
2:   Select the protective equipment to restock & Users   need to compare the numerical data in the table, discard items according to   their individual information and choose the one with the lowest number.  & Table with icons and numerical data, Text & CEO   \\
3:   Select one offer in the overview  & Users   first need to filter offers according to the personal information provided   (Task 3.1) and then select the cheapest offer that fulfils all criteria (Task   3.2).  & Scatterplot,   Text & Head   of ICU, Head of finance, Head of logistics, CEO  \\
4:   Select delivery option  & Users need to   go through the delivery options and select the one that fulfils all criteria.   To do so, they need to read the details and check the trajectory on the map. & Table   with icons, textual and numerical data, Map, Text & Head of finance,
Head of logistics  \\ \bottomrule
\end{tabular}  
\end{table*}

\section{User study}

We conducted an observational user study with 24 persons (six groups of four participants) that we asked to solve four tasks on an interactive wall display. While the user study was conducted to analyse a broader set of behavioural patterns (including head direction, hand gestures and body movements), in this paper we focus on mid-air pointing gestures and more specifically on the following research question: Which types of mid-air pointing gestures are made by groups during collaborative work in a wall-display environment and how do they contribute to awareness information? We apply a qualitative approach based on the analysis of audio-visual data to obtain findings about the type and frequencies of users’ pointing gestures.

\subsection{The decision-making scenario}

The decision-making scenario deals with the analysis of data from different types of sources and is therefore one of the most common types of wall-display applications that were identified by \cite{beaudouin2012multisurface}. It was developed based on a previous project that aimed at creating a control tower for experts and decision-makers to help managing medical supply chains during distress times (e.g., a pandemic). The scenario was adapted for the context of the user study, in particular to i) show different types of data, ii) integrate collaborative mechanisms, and iii) be solvable by non-experts (e.g., students).
The scenario tackles the problem experienced during the first wave of COVID-19, where hospitals in Luxembourg had difficulties obtaining the needed personal protective equipment (e.g., surgical masks, goggles, gowns) on time. The usual intermediary suppliers were no longer delivering, and so the hospitals had to take care of all the tasks related to ordering new stock. The study participants act as a team of decision-makers from a hospital that needs to ensure that the stock of protective equipment meets the hospital’s needs for the next three months. More specifically, they need to i) estimate the upcoming new COVID-19 infections and hospital occupancy, ii) select one type of equipment where the strongest stock shortage is to be expected, iii) select an offer to replenish the stock, and iv) select a delivery method. For each of the steps different types of data is provided (see Table 1) and juxtaposed onto the wall display. Participants need to interpret and analyse the data, and identify the best solution given the existing constraints.

In addition to the different types of data, we also integrated mechanisms to make the tasks collaborative. Building on principles of positive interdependency \cite{johnson1999making}, we distributed the required resources among participants, so that each of them had unique and complementary competencies that were needed to solve the tasks. In particular, we made use of four role cards, each indicating a profile (CEO, head of ICU, head of finance, and head of logistics) as well as information and objectives specifc to that profile (e.g., one knows that they will receive new FFP2 masks by the minister, another will only accept the cheapest option). That way, the information is distributed among participants and only when they work together, they are able to fend the best solution to the task. In general, our scenario supports the four broad categories of behaviours relevant to decision-making activities based on the literature (see \cite{tong2017supporting}).
The scenario was implemented using DeBORAh \cite{vandenabeele2022deborah}, a front-end, web-based software layer supporting the orchestration of interactive spaces combining multiple devices and screens. DeBORAh uses a flexible approach to decompose multimedia content into containers and webpages, so that they can be displayed in various ways across devices. The different views have been implemented as webpages using WebSocket to communicate with the server-side Node-RED installation which handles the workflow and logic part of the scenario. The system supports touch input on its screens, as well as using up to two HTC Vive Controllers to point and interact with the scenario from the distance. Available actions include modifying values, panning, and zooming the map, and selecting data. The Vive Controllers enable in addition to display a coloured pointer on the screen.

\subsection{Participants}
24 persons, divided into six groups of four, participated in our study. 20 of the participants were male; there were three females, and one participant preferred not to answer the question. Participants’ ages ranged from 20 to 38 (M=24.2, SD=4.5, Mdn=22). All participants were students in either computer science or geography. We asked the participants about their familiarity with the teammates in a form of a five-point Likert scale from (1) “not at all familiar” to (5) “very familiar” (other scale ticks were unlabelled). Most participants were at least somewhat familiar with others from the same group. In fact, 50 out of the total 72 peer-to-peer relations were labelled as
(4) or (5), with only 10 on (1) and 6 for both (2) and (3).
Since the scenario involved a large display setup, participants were asked about the frequency of their usage of large interactive displays. Of the 24 participants, 10 reported no prior experience with such displays, while the other 14 indicated having some experience, with 6 cases of yearly, 4 of monthly and 3 of daily use being reported. Similarly, because HTC Vive controllers were offered as an input option, we asked participants how often they use similar pointing devices (e.g., virtual reality controllers, Wiimote), to which most replied that they had some prior experience (only 3 said they had never used such devices before, and 14 cases of yearly, 3 of monthly, 3 of weekly, and 1 of daily use mentioned). During the study, the HTC Vive controllers were used only by one group. In the debriefing, participants mentioned that there was no particular need for using the controllers as the available touch interaction was working well, and as they were sufficiently close to the screens.

\subsection{Setup and procedure}
The experiment took place in October 2022 in our 360° immersive arena, 2m high, composed of 12 screens (4K resolution each) that are spatially positioned in a circle of 3.64m diameter (see Figure \ref{fig:2}). Eight of the screens were used as surfaces for the study.
In consultation with our Ethics committee, we first informed participants of the objectives and context of the study.We explained to them what type of data we would record and how we would store and process it. To respect privacy, participants would immediately be assigned pseudonyms, and all documentation would be organized with these. We also informed participants about their rights to withdraw their consent and ask for the deletion of the data at any time and without giving reasons. We provided them with an information sheet and a consent form to sign.
Participants were then instructed to pick a role card and to only mention the information to the others when negotiating in these tasks. They were then led to the experimentation room to participate in a 5-minutes demonstration presenting the types of input they had access to (touchscreen and HTC Vive controllers) and the kind of interface elements they would need to manipulate (buttons, sliders, 3D globe). The experimenter (part of the research team) gave the participants an introduction to the scenario, together with general instructions on how to proceed through the tasks. Once the introduction was finished, the experimenter left the room and let participants begin the prescribed tasks. After all the tasks had been completed, the participants were provided with a questionnaire, the analysis of which, however, is out of scope of this paper.

\subsection{Data collection and analysis}
\begin{figure*}
    \centering
  \includegraphics[width=.8\textwidth]{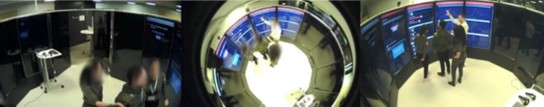}
  \caption{Three different camera angles (top, front and back) were recorded by fxed cameras in the lab and used for the annotation.}
  \Description{}
  \label{fig:2}
\end{figure*}

The primary source of data was audio-visual recordings of the study. We recorded the decision-making process using three fixed cameras (top, front, and back cameras), see Figure \ref{fig:2}.
The audio-visual data was analysed manually using ELAN \cite{wittenburg2006elan} by one researcher with extensive experience in gesture analysis. The annotation of the type of gesture a user performed in the exact time frame in which it occurred was done for all participants. The various camera angles helped to annotate properly the gesture performance, since in some angles, due to the users’ body position and occlusions by other participants, the hand gesture was not clearly visible.
As mentioned before, mid-air pointing gestures have been explored mostly on tabletops (e.g.,\cite{anastasiou2014gesture, banerjee2011pointable, hinrichs2011gestures, soni2021affording}). Based both on the literature and analysis of the video footage obtained, we developed our own gesture taxonomy. Since this gesture taxonomy is the first attempt towards a taxonomy of mid-air pointing gestures on wall displays, the annotator ran two rounds of annotation. During the first round, the subtype and duration were annotated, and in the second round, the focus was on the referent and the target.

\section{Results}
\begin{figure*}
    \centering
  \includegraphics[width=.8\textwidth]{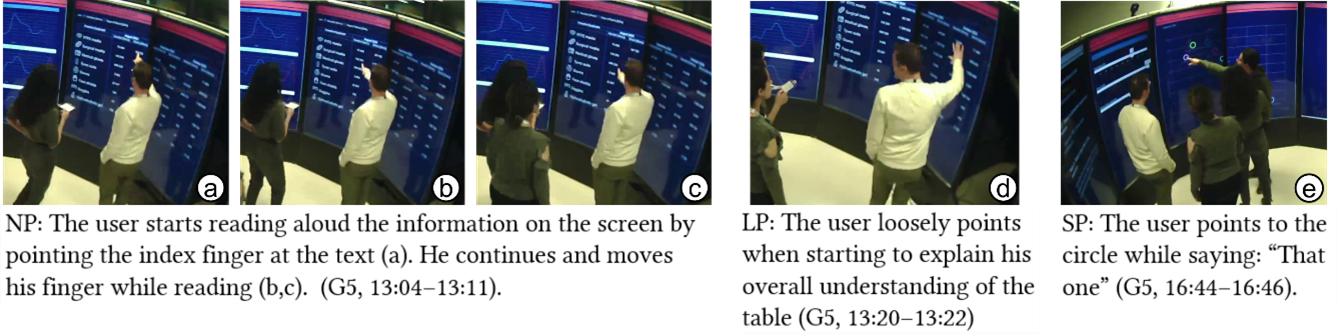}
  \caption{Example occurrences of the three types of mid-air pointing gestures.}
  \Description{}
  \label{fig:3}
\end{figure*}

In our analysis, we observed that users’ gestures showed different characteristics (hand usage, distance from screen, referent) depending on the communicative function of the gesture. Therefore, we created a gesture typology based both on the external characteristics of the gesture and their communicative function. This typology is usable by other researchers who analyse mid-air pointing gestures on large displays. We have excluded in our results touch-pointing gestures as these were done to interact with the application and thus outside our focus to study the support for collaboration.
A first type of mid-air pointing gesture we identified was narrative pointing (NP). During NP, a user points sharply with the index fnger, but also moves the finger towards a larger area of the display (up/down or left/right) (see Figure \ref{fig:3}a-c). This gesture is speech-accompanied: the user either reads the text displayed or explains a graph (increase/decrease of values). It is also longer in duration than the other two subtypes. A second type of mid-air pointing gesture we identified is loose pointing (LP). Here, the user is not looking at the screen and holds the hand usually open (see Figure \ref{fig:3}d) or palm up. LP gestures have vague communicative goals, are usually not speech-accompanied and are of very small duration. They often happen when a user describes a concept as a whole. They are not part of a sequence where the users touch the display. Finally, we identified a third category of sharp pointing (SP). This is the “stereotypical” pointing gesture with an index finger where a user points to a very specific area of the display (e.g., a specific word or number) (see Figure \ref{fig:3}e). Its duration is usually much shorter than narrative pointing. Its origin comes from children when making requests (e.g. \cite{o1996two}).
Noteworthy is the distinction between sharp and narrative pointing. The narrative pointing is not a succession of sharp pointing gestures since it does not include retraction of hands in between. A narrative pointing is a single, continuous, long-in-duration gesture, whereby the user explains or describes a concept. In contrast, the function of sharp pointing is to quickly point to a specific value or word at the screen in order to support his/her point through evidence. The narrative pointing vs. sharp pointing is analogous to the work of Okuya et al. \cite{okuya2020investigating}, where, in a scenario about CAD data adjustment, they distinguished between deictic-specific vs. deictic-range with regards to a specific shape or range of shapes accordingly. However, in our decision-making scenario, we do not limit NP into ranges, since a NP can also include reading a tweet from the news or describe on the map the travel of a delivery option, for instance.

\begin{table*}
  \caption{Subtypes of pointing gestures on large displays with their characteristics.}
  \label{tab:2}
 \begin{tabular}{p{3cm}p{2.9cm}p{4cm}p{3.5cm}p{2cm}}
 \toprule
Type                                                                                  & Hand Usage                                               & Referent                                                     & Target                                    & Duration \\ \midrule
Narrative pointing (NP) (Figure \ref{fig:3}a-c) & Pointing and moving index finger up/down or   left/right & Full sentence / description of a value increase /   decrease & A specific long area at the screen        & Long     \\
Loose pointing (LP) (Figure \ref{fig:3}d)                                                       & Open palm, two fingers                                   & Concept in total                                             & A bit far of the screen                   & Mid      \\
Sharp pointing (SP) (Figure \ref{fig:3}e)                                                       & Index finger                                             & Specific value / text                                        & A specific small area/point at the screen & Short     
\\ \bottomrule
\end{tabular}  
\end{table*}

\begin{figure}
    \centering
  \includegraphics[width=.35\textwidth]{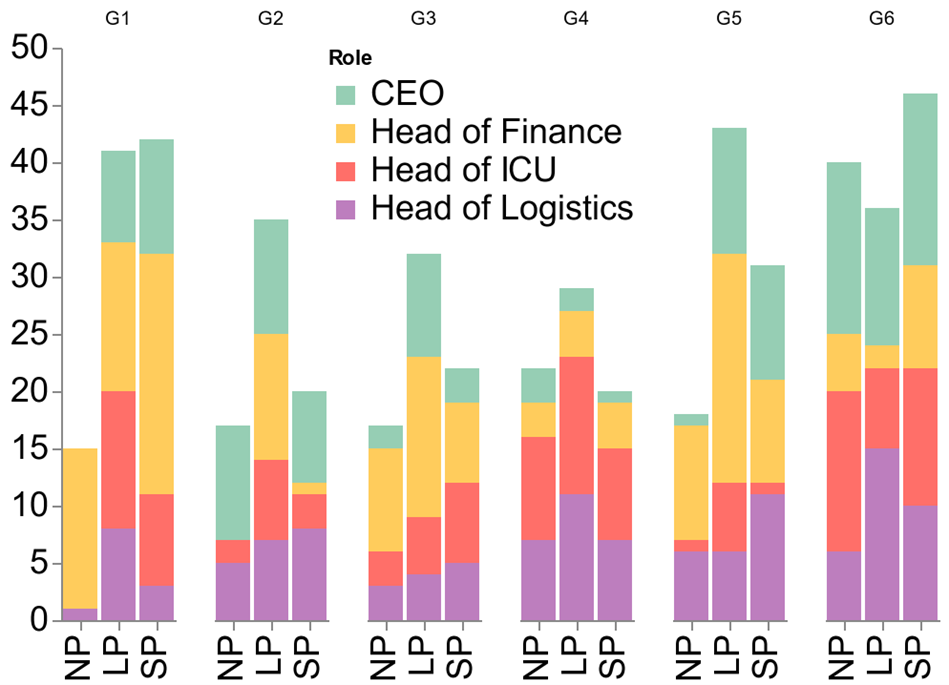}
  \caption{Bar chart with the number pointing gestures per group, subtype and role.}
  \Description{}
  \label{fig:4}
\end{figure}

The pointing gestures of each category therefore differed with regard to hand/palm usage, referent, target and duration (see Table 2). The referent is the intended meaning (concept to be communicated) of the pointing gesture (e.g., a narrative pointing can refer to a line or an (imaginary) movement, or an outline of a shape) and the target is the specific physical location, in our specific case, a specific area of the display. It should be noted that these three types are all mid-air gestures, i.e., they are contact-free, i.e., the users did not touch the display. If a user touches the display, we regard this as a sequence of gestures in the form of pointing-touch; these gestures are outside the scope of this paper because of space constraints, but are part of future work.
To investigate the frequencies of the pointing gestures, we counted the frequencies by subtype, group and user (see Table 3). We can deduce that all three subtypes are commonly used by all groups: SP and LP were used by all users, and NP by 21 users out of 24. Figure \ref{fig:4} shows the number of pointing gestures of each subtype per group and role. NP occurred on average 6 times per group (SD=2.11), LP 9.1 times (SD=1.3), and SP 7.63 times (SD=3.02). In 4 out of 6 groups, LP is most frequent. This shows that users use often pointing gestures to express their thoughts and not necessarily for showing to others a specific area of the display. LP is analogous to consequential communication \cite{segal1994effects}, where the producer of the information does not intentionally undertake actions to inform the other person, however, these gestures provide a great deal of information. It has also been shown that pointing gestures in problem-solving tasks on tangible interfaces encourage the use of rapid epistemic actions (a common and shared meaning for both the gesturer and the other group members \cite{anastasiou2014gesture}). Another reason for the high usage of LP might also be the characteristics of the wall display environment in this study. Because of the larger distance to the screen and the scaling up of the data, users might perceive approximative pointing as sufficient and do not need to point at a specific area.

\begin{table*}[]
\begin{tabular}{lllllllllllllllllll}
\toprule
                  & \multicolumn{3}{l}{Group 1} & \multicolumn{3}{l}{Group 2} & \multicolumn{3}{l}{Group 3} & \multicolumn{3}{l}{Group 4} & \multicolumn{3}{l}{Group 5} & \multicolumn{3}{l}{Group 6} \\ \cline{2-19}
                  \midrule
                  & NP              & LP              & SP              & NP              & LP              & SP              & NP              & LP              & SP              & NP              & LP              & SP              & NP              & LP              & SP              & NP              & LP              & SP              \\
Head of ICU       & 0               & 12              & 8               & 2               & 7               & 3               & 3               & 5               & 7               & 9               & 12              & 8               & 1               & 6               & 1               & 14              & 7               & 12              \\
Head of Finance   & 14              & 13              & 21              & 0               & 11              & 1               & 9               & 14              & 7               & 3               & 4               & 4               & 10              & 20              & 9               & 5               & 2               & 9               \\
Head of Logistics & 1               & 8               & 3               & 5               & 7               & 8               & 3               & 4               & 5               & 7               & 11              & 7               & 6               & 6               & 11              & 6               & 15              & 10              \\
CEO               & 0               & 8               & 10              & 10              & 10              & 8               & 2               & 9               & 3               & 3               & 2               & 1               & 1               & 11              & 10              & 15              & 12              & 15  \\ \bottomrule
\end{tabular}
\end{table*}

\section{Discussion}
As mentioned previously, this collocated study was conducted with the aim to observe the most relevant non-verbal cues to consider when designing awareness support for remote collaboration. In previous work it was already observed that in front of a large screen, users are often moving and making use of hand gestures (e.g., \cite{langner2018multiple}). Our results confirm this observation and show that three types pointing gestures play an important role for establishing awareness during collocated decision-making at wall displays: i) narrative pointing, ii) loose pointing, and iii) sharp pointing. They show different external characteristics and are made for different communicative purposes, therefore, they need adjusted designs when reproduced remotely as an awareness cue. More specifically, whereas sharp pointing should be translated into a smaller sized shape indicating a precise location, loose pointing should highlight a larger area, such as an entire screen (e.g., using a coloured border). The visualisation of narrative pointing can be less precise than sharp pointing but should be shown for a longer duration to accompany the flow of the narration. On a more general note, we may also include avatars that implement these three types of gestures for providing feedback or assisting users in their decision-making in a more natural way. This requires working on user tracking solutions to gather the necessary data, whose interpretation will then be conveyed to remote users.

\section{Conclusion and future work}
Co-speech gestures are a very intuitive and expressive means of communication and therefore play a crucial role in collaborative decision-making. In this paper we described a user study with 24 participants, who, split into groups of four, had to collaboratively fulfil tasks related to the organization of supply chains of medical equipment during a new COVID-19 wave. We investigated mid-air pointing gestures and their subtypes based on their communicative function as awareness cues. Based on the mean of each subtype of the pointing gestures, the loose pointing was most frequently used, followed by sharp pointing and narrative pointing. The reliance on gesture subtypes varied across participants and groups, and sometimes vague pointing was sufficient to support verbal negotiations. The proposed gesture types can be reused in other gesture analyses and further explored for different contexts or setups, including remote or mixed presence collaboration.
As part of future work, a deeper dive is needed into factors influencing the reliance on gestures we identified, for example to evaluate whether the type of task, role or prior familiarity of users, task completion time or even data visualization on the displays makes a difference on which type and how many gestures are performed. Furthermore, we focused so far on certain specific gestures. In our study, users performed also other types of gestures, such as emblems and adaptors (see \cite{ekman1972hand}), and iconic gestures (see \cite{mcneill2005gesture}), which will need to be analysed in the near future. For a full analysis, we will include multiple annotators and calculate the inter-annotator agreement with Kohen’s Kappa. We also plan on performing speech transcriptions and speech-gesture alignments. Moreover, we will work on identifying sequences of gestures and which pointing gestures resulted in touching the display.

\begin{acks}
Authors would like to thank all participants of the user study as well as the Luxembourg National Research Fund (FNR) for funding this research work under the FNR CORE ReSurf project (Grant nr C21/IS/15883550).
\end{acks}

\bibliographystyle{ieeetr}
\bibliography{sources}


\end{document}